\documentclass[preprint,12pt,sort&compress]{elsarticle}

\usepackage{graphicx}% Include figure files
\usepackage{dcolumn}% Align table columns on decimal point
\usepackage{bm}% bold math
\usepackage{amsmath}
\usepackage{subfigure}
\usepackage{color}
\newcommand{\f}{\frac}

\newcommand{\eps}{\varepsilon}

\usepackage{amssymb}

\journal{Frontiers of Physics}

\begin{document}

\begin{frontmatter}

\title{Extended social force model with a dynamic navigation field for bidirectional pedestrian flow}

\author[label1]{Yan-Qun Jiang\corref{cor1}\fnref{fn1}}
\ead{jyq2005@mail.ustc.edu.cn}
\author[label2]{Bo-Kui Chen\corref{cor1}\fnref{fn1}}
\ead{chenssx@mail.ustc.edu.cn}
\author[label3]{Bing-Hong Wang\corref{cor1}}
\ead{bhwang@ustc.edu.cn}
\author[label2]{Weng-Fai Wong}
\author[label4]{Bing-Yang Cao}
\fntext[fn1]{First author: These authors contributed equally to this
work.} \cortext[cor1]{Corresponding author.}
\address[label1]{\fnref{label1} School of Science, Southwest University of Science and Technology, Mianyang, Sichuan 621000, China}
\address[label2]{\fnref{label2} School of Computing, National University of Singapore, Singapore 117417, Singapore}
\address[label3]{\fnref{label3} Department of Modern Physics and Nonlinear Science Center, University of Science and Technology of China, Hefei, Anhui 230026, China}
\address[label4]{\fnref{label4} Key Laboratory for Thermal Science and Power Engineering
of Ministry of Education, Department of Engineering Mechanics,
Tsinghua University, Beijing 100084, China}

\begin{abstract}
An extended social force model with a dynamic
navigation field is proposed to study bidirectional pedestrian
movement. The dynamic navigation field is introduced to describe the
desired direction of pedestrian motion resulting from the
decision-making processes of pedestrians. The macroscopic fundamental
diagrams obtained using the extended model are validated against camera-based
observations. Numerical results show that this extended
model can reproduce collective phenomena in pedestrian traffic, such
as dynamic multilane flow and stable separate-lane flow.
Pedestrians' path choice behavior significantly affects the
probability of congestion and the number of self-organized lanes.
\end{abstract}


\end{frontmatter}

% \linenumbers


\section{Introduction}
\label{sec:1} Pedestrian movement is an important factor in the
design and optimization of transportation facilities, pedestrian
walkways, and public transport intersections. Further, complex situations
involving bidirectional pedestrian movement occur continually when
enormous numbers of people with different final destinations use
pedestrian facilities, such as crosswalks, sidewalks, corridors, and
stairways. Therefore, problems related to bidirectional pedestrian
flow and its effects on pedestrian dynamics have
attracted considerable attention from scientists and engineers in recent decades
\cite{Helbing1995,Blue2001,Lam2002,Hughes2002,Isobe2004,Lakoba2005,Wong2010}.

Many behavioral investigations and empirical studies
\cite{Lam2002,Wong2010,Helbing2005,Kretz2006,Zhang2012,Zhang2014316,Saberi2015120}
have been conducted to gain a good understanding of bidirectional
pedestrian flow characteristics.  Many phenomena such as
self-organization and synchronization have been observed in
pedestrian flow \cite{Helbing2005,Helbing1999,m2011}.
Synchronization has also been studied in other fields
\cite{wang2017,ma2015}. Empirical observations are preferable for
analyzing qualitative self-organization phenomena, such as the
formation of lanes and the occurrence of congestion at sufficiently
high densities, and obtaining fundamental diagrams. The fundamental
diagrams representing the relationships among density, velocity, and
flow generally differ for various walking facilities
\cite{Zhang2012,Zhang2014316}. They can quantify the capacity of
pedestrian facilities and support the construction of reliable
pedestrian simulation models
\cite{Lam2002,Isobe2004,Zanlungo2011,FlOtterOd2015194}. A number of
simulation models for pedestrian dynamics, including hydrodynamic
models
\cite{Hughes2002,Xiong2011,Jiang2012,Hoogendoorn2014684,Jiang201569,Jiang201536},
mesoscopic (kinetic) models \cite{Bellomo2011,Bellomo2015a}, the
social force model
\cite{Helbing1995,Helbing2000,Yang2014456,Guo2014428, Hou2014,
Korecki2016}, the cellular automaton model
\cite{Blue2001,Weng2006,Jian2014}, and the lattice gas model
\cite{Tajima2002,Nagai2006503}, have been proposed recently to
simulate and reproduce collective behaviors appearing in
bidirectional pedestrian movement. Examples of collective patterns
caused by local interactions among conflicting pedestrians are
jamming, lane formation, and oscillations at narrow bottlenecks in
bidirectional pedestrian streams
\cite{Guo2014428,Tajima2002,Fu20161}. Optimal self-organized
phenomena such as the spontaneous formation of lanes of uniform
walking direction can help to reduce conflicts with pedestrians
moving in the opposite direction and increase the efficiency of
walking \cite{Helbing2005,Helbing1999}.

Most simulation models of bidirectional pedestrian flow can describe the basic behavioral characteristics qualitatively
\cite{Helbing1995,Blue2001,Hughes2002,Xiong2011,Jiang2012,Hoogendoorn2014684,Jiang201569,
Yang2014456,Guo2014428,Weng2006,Jian2014,Tajima2002,Nagai2006503}.
In this work, we consider only the social force model owing to its many
advantages. For example, it captures the mutual
influence of individual pedestrians in a two-dimensional (2D)
continuous space by defining social analogs of physical forces,
e.g., repulsive interactions, frictional forces, dissipation, and
fluctuations \cite{Helbing1995}. In this way, the detailed behavior of
individual pedestrians can be considered, which makes the
simulated pedestrians more realistic \cite{Yang2014456}. The social
force model first proposed by Helbing and Moln\'{a}r
\cite{Helbing1995} has been modified in numerous ways
\cite{Lakoba2005,Guo2014428,Parisi2007343,Parisi20093600,Kwak2013,
Kretz2011,Karamouzas2014}. For instance, Guo \cite{Guo2014428}
introduced spatial and temporal separation rules to reproduce
self-organizing movement patterns of pedestrians, e.g., oscillatory
flow and three classes of lane formation: unidirectional,
mixed, and separate. Kretz \textit{et al.} introduced the
notion of the ``quickest path'' using a non-iterative method to
describe the path choice behavior of pedestrians \cite{Kretz2011}.
Karamouzas \textit{et al.} used a novel statistical--mechanical
approach to describe human interactions across a wide variety of
situations, speeds, and densities \cite{Karamouzas2014}. However, in the
social force model, where pedestrians choose their paths by a shortest-distance strategy, the decision-making process of pedestrians walking in a
dynamic environment is tactically simplified. Path choice strategy is not only a research
field in pedestrian flow, but also a hot topic in traffic flow, and
many research results have been reported
\cite{Wahle2000,Wang2005,chen2011,chen2012a,chen2012b,chen2012c,chen2016}.

In this paper, we revise Helbing's model by introducing a
reactive user-equilibrium path choice strategy to simulate
bidirectional pedestrian flow and analyze the effect of path choice
behaviors on bidirectional pedestrian movement. This path choice
strategy, which determines a continuous dynamic navigation field,
i.e., the desired walking direction of individual pedestrians, has
been used in some hydrodynamic models
\cite{Xiong2011,Jiang2012,Jiang201569}, in which
pedestrians in a group, acting as a ``thinking'' crowd flow, are assumed to choose a
path with the lowest instantaneous walking cost on the basis of
the instantaneous traffic information available to them when they
make their decisions. These hydrodynamic models can describe well
the global properties of the pedestrians but cannot describe in detail
the behavioral characteristics of individual pedestrians, such as their
static and dynamic obstacle avoidance behaviors. The paper is structured as follows. The extended social force model of
bidirectional pedestrian traffic and its numerical method are
described in Section 2. Numerical simulation results are presented
in Section 3. Finally, Section 4 summarizes the results and presents
concluding remarks.

\section{Dynamic pedestrian model}
\label{sec:2} The well-known social force model \cite{Helbing1995}
is a microscopic, force-based model in which all pedestrians are
treated as self-driven particles, and their behaviors are described
by a mixture of socio-psychological and physical forces. This model
is developed to simulate two groups of pedestrians, Group 1
and Group 2, walking in a 2D continuous walking
facility denoted as $\Omega$. The pedestrians in each group have the
same travel destination, e.g., one of the two exits of the walking
facility. For instance, each pedestrian in Group $k~(k\in\{1,2\})$,
who has mass $m^{k}_i$ and radius $r^{k}_i$ and is located at the
position $\mathbf{r}^k_i(t)\in \Omega$, prefers to move with a
certain desired speed $v^k_d$ in a certain direction
$\mathbf{e}^k_i(t)$, and thus tends to adapt his or
her actual velocity $\mathbf{v}^k_i(t)$ accordingly with a certain
characteristic time $\tau^k_i$. Simultaneously, pedestrian $i$ in
Group $k$ will try to maintain a distance from another pedestrian $j$ in
the two groups and from the walls $w$ according to the interaction forces
$\mathbf{f}^k_{ij}$ and $\mathbf{f}^k_{iw}$, respectively. To take into account the effect of perception, we define the
sensory field $R^k_i$ for pedestrian $i$ in Group $k$ as
\begin{equation}\label{1}
R^k_i=\{\mathbf{r}\in \Omega:  (\mathbf{r}-\mathbf{r}^k_i)\cdot
\mathbf{e}^k_i > 0, |\mathbf{r}-\mathbf{r}^k_i|\le R_0 \},
\end{equation}
where $ R_0$ represents the sensory range of pedestrians. The
sensory field $R^k_i$ is usually anisotropic and short-range, which
means that pedestrian $i$ in Group $k$ can interact only with
pedestrians who are within his/her sensory field.

\subsection{\label{sec:2.1}Model description}
The movement of individual pedestrians in Group $k$ can be described
by nonlinear ordinary differential equations (ODEs)
according to Newton's law:
\begin{eqnarray}\label{2}
\f{d \mathbf{r}^k_i}{d t}
&=&min(v_{max},\|\mathbf{v}^k_i\|)\f{\mathbf{v}^k_i}{\|\mathbf{v}^k_i\|},
\\ \label{3}
\f{d \mathbf{v}^k_i}{d t}
&=&\f{v^k_d\mathbf{e}^k_i-\mathbf{v}^k_i}{\tau^k_i}+\f{\sum_{j\ne
i}\mathbf{f}^k_{ij}+\sum_{w}\mathbf{f}^k_{iw}}{m^k_i}.
\end{eqnarray}
Here, the change in position $\mathbf{r}^k_i(t)$ is given by the
velocity $\mathbf{v}^k_i(t)$ of pedestrian $i$ in Group $k$, and
$v_{max}$ is the maximal acceptable speed of pedestrians.

In Eq. (\ref{3}), the first force,
$(v_d^k\mathbf{e}^k_i-\mathbf{v}^k_i)/\tau^k_i$, is an attractive
force that drives pedestrian $i$ in Group $k$ to move toward
his/her goal. The interaction force $\mathbf{f}^k_{ij}$ includes
the social force $\mathbf{F}^k_{ij}$ and the granular force
$\mathbf{G}^k_{ij}$; i.e.,
$\mathbf{f}^k_{ij}=\mathbf{F}^k_{ij}+\mathbf{G}^k_{ij}$. Here, the
social force $\mathbf{F}^k_{ij}$ is defined as
\begin{equation}\label{4}
\mathbf{F}^k_{ij}=A\text{exp}[(r_{ij}-d_{ij})/B]cos(\theta_{ij})\mathbf{n}_{ij},
j\in M_i,
\end{equation}
where $A$ and $B$ are constants that determine the strength and
range of the interaction, respectively; $r_{ij}$ is the sum of the radii of
pedestrians $i$ and $j$; $d_{ij}$ denotes the distance between
pedestrians $i$ and $j$; $\mathbf{n}_{ij}$ is the unit vector
pointing from pedestrian $j$ to pedestrian $i$; $\theta_{ij}$ is
the angle between $\mathbf{e}^k_i$ and $-\mathbf{n}_{ij}$; and
$M_i$ is the set of all pedestrians in the sensory region $R^k_i$.
The granular force $\mathbf{G}^k_{ij}$ is described as
\begin{equation}\label{5}
\mathbf{G}^k_{ij}=-\eps_{ij}\gamma\mathbf{n}_{ij}-\eps_{ij}\kappa
\Delta v^t_{ji}\mathbf{t}_{ij}+\eta  \Delta v^n_{ji}\mathbf{n}_{ij},
j\in N_i,
\end{equation}
where $\eps_{ij}=r_{ij}-d_{ij}$; $\gamma$, $\kappa$, and $\eta$
determine obstruction effects in physical interactions;
$\Delta v^t_{ji}$ and $\Delta v^n_{ji}$ are the tangential and
normal velocity differences, respectively; $\mathbf{t}_{ij}$ is the
tangential unit vector, which is orthogonal to $\mathbf{n}_{ij}$;
and $N_i$ is the set of all pedestrians who have any physical
contact with pedestrian $i$ in Group $k$, i.e., $\eps_{ij}\le0$. In
Eq. (\ref{5}), the term $\eta  \Delta v^n_{ji}\mathbf{n}_{ij}$
represents a physical damping force with the damping parameter
$\eta$ \cite{Parisi2007343}.

The interaction force $\mathbf{f}^k_{iw}$
generated by the interaction between pedestrian $i$ in Group $k$ and
the nearest wall $w$ such that the pedestrian avoids the wall before touching it is
described in a form analogous to that of the two-body interaction
$\mathbf{f}^k_{ij}$.

\subsection{Definition of the dynamic navigation field}
We assume that all pedestrians in Group $k$ are familiar with the
surroundings, e.g., the locations of walls and exits, and know the
current traffic conditions that obtain when they make
decisions. They prefer to walk in a reactive user-optimal manner to
minimize their total instantaneous cost from location $\mathbf{r}$
to the final destination, e.g., one of the two exits of the walking
facility. For Group $k$, the minimum total instantaneous walking
cost is denoted as $\Phi^k(\mathbf{r})$, and $\Phi^k(\mathbf{r})=0$
at the exit represented by $\Gamma^k$.

We denote the local travel cost at location $\mathbf{r}$ as
$C^k(\mathbf{r})$; it represents the walking cost per unit
distance of movement incurred by pedestrians in Group $k$ and is
defined as
\begin{equation}\label{6}
C^k(\mathbf{r})=\frac{1}{v^k_d}+\omega_1(1-cos(\Psi))(\rho^{l(\ne
k)}(\mathbf{r}))^2+\omega_2(\rho^1(\mathbf{r})+\rho^2(\mathbf{r}))^2.
\end{equation}
Here, $l\in\{1,2\}$; $\omega_1$ and $\omega_2$ are weight
coefficients; $\Psi$ is the angle between the vectors
$\mathbf{e}^1(\mathbf{r})$ and $\mathbf{e}^2(\mathbf{r})$, where
$\mathbf{e}^k(\mathbf{r})$ is the desired direction of motion for
Group $k$ at location $\mathbf{r}$; and $\rho^k(\mathbf{r})$ is the
local density of Group $k$ at location $\mathbf{r}$. In Eq.
(\ref{6}), the term $1/v_d^k$, which is the major factor, represents the
cost associated with the travel time, and the terms
$\omega_1(1-cos(\Psi))(\rho^{l(\ne k)}(\mathbf{r}))^2$ and
$\omega_2(\rho(\mathbf{r}))^2$, which are the minor factors, represent
other associated costs such as a preference for reducing collision
conflicts with pedestrians in the other group and avoiding
high-density regions to feel comfortable, respectively. For Group
$k$, the local density $\rho^k(\mathbf{r})$, which is used to
evaluate the crowdedness level within the subarea at location
$\mathbf{r}(t)$, is measured as
\begin{equation}\label{7}
 \rho^k(\mathbf{r})=\sum_i f(\|\mathbf{r}^k_i-\mathbf{r}\|),
\end{equation}
where
\begin{equation}
f(z)=\frac{\exp(\frac{-z^2}{2R^2})}{2\pi R^2}
\end{equation}
with a measurement parameter $R=0.7$~m.

The minimum total instantaneous walking cost $\Phi^k(\mathbf{r})$
incurred by pedestrians in Group $k$ from the origin $\mathbf{r}$ to
the destination $\Gamma^k$ is calculated as
\begin{equation}\label{8}
\Phi^k(\mathbf{r})=\Phi^k(\mathbf{r}_d)+\underset{p}{min} \int_p
C^k(\mathbf{r})ds,
\end{equation}
where $\mathbf{r}_d\in\Gamma^k$, and $p$ denotes any path from the origin
$\mathbf{r}$ to the destination $\Gamma^k$. According to \cite{Jiang2012},
the rate of reduction in the cost potential $\Phi^k(\mathbf{r})$ for
Group $k$ is greatest along the direction of motion. Thus, the
desired direction of motion for Group $k$ at location $\mathbf{r}$
is obtained as \cite{Jiang2012,Jiang201569}
\begin{equation}\label{9}
\mathbf{e}^k(\mathbf{r})=-\f{\nabla\Phi^k(\mathbf{r})}{\|\nabla\Phi^k(\mathbf{r})\|},
\end{equation}
which determines a dynamic navigation field for bidirectional
pedestrian movement in the 2D continuous domain $\Omega$.

Furthermore, the cost potential for Group $k$ satisfies the following
eikonal equation \cite{Jiang2012}:
\begin{eqnarray}\label{10}
\| \nabla\Phi^k(\mathbf{r})\|&=&C^k(\mathbf{r}), \mathbf{r}\in
\Omega,
\\\nonumber\Phi^k(\mathbf{r})&=&0, ~~~~~~\mathbf{r}\in \Gamma^k.
\end{eqnarray}

The extended social force model described
in Section 2 is solved as follows.
\begin{enumerate}[ Step 1.]
\item  Given the location and speed of pedestrian $i$ in Group $k$,
$\mathbf{r}^{k,n}_i$ and $\mathbf{v}^{k,n}_i$, respectively, at time
$t^n~(n=0,1,2,...)$, estimate the local density $\rho^k(\mathbf{r})$
using Eq. (\ref{7}), and then solve Eq. (\ref{10}) by a fast
sweeping method \cite{Jiang2012,Zhang2006} to determine the dynamic
navigation field for bidirectional pedestrian flow. The desired
walking direction of pedestrian $i$ in Eq. (\ref{3}) is obtained using
Eq. (\ref{9}).

\item Solve the system of first-order ODEs composed of Eqs.
(\ref{2}) and (\ref{3}) by the second-order Runge--Kutta method and
then update its solution as $\mathbf{r}^{k,n+1}_i$ and
$\mathbf{v}^{k,n+1}_i$ at time $t^{n+1}=t^n+\Delta t$.

\item Stop the calculation process if all pedestrians leave the walking facility.
Otherwise, replace $n$ with $n + 1$ and go to Step 1.
\end{enumerate}
Here, $\Delta t$ is the size of the time step in the simulation.

\section{Numerical simulations}
Several parameters in our model are set as follows: for each pedestrian in
Group $k$, $r^k_i=0.25$~m, $m^k_i=80$~kg, $\tau^k_i=0.5$~s,
$A=2000$~N, $B=0.08$~m, $k=1.2\times10^5$~kg/s$^2$,
$\kappa=2.4\times10^5$~kg/m/s, and $\eta=100$~kg/s
\cite{Helbing2000,Parisi2007343}. The sensory radius in Eq.
(\ref{1}) is $R_0=3$~m. In Eq. (\ref{2}), the desired walking speed
is $v^k_d=1.034$~m/s \cite{Wong2010}, and the maximum walking speed
is $v_{max}=1.3v_f$. In Eq. (\ref{6}), $\omega_2=0.05$. The size of
the time step in the simulation, $\Delta t$, is set to $0.01$~s.

\begin{figure}
\centering
\includegraphics[width=\textwidth]{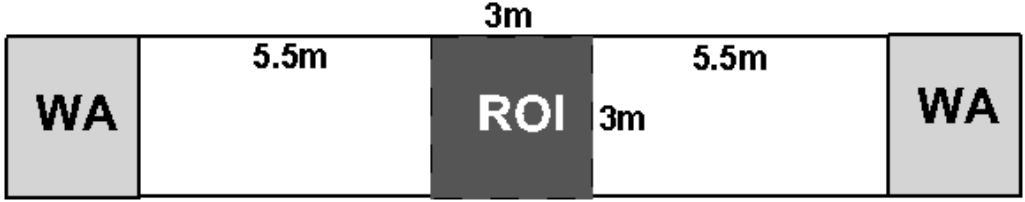}
\caption{Walkway configuration for the simulation. WA: Waiting area;
ROI: Region of interest.}\label{fig1}
\end{figure}

To calibrate and test the extended model, we used experimental data
from a series of controlled experiments of bidirectional pedestrian
streams collected by Wong \textit{et al.} \cite{Wong2010}. The
macroscopic fundamental diagrams of bidirectional pedestrian flow
were validated against camera-based observations. Numerical
simulations of bidirectional pedestrian movement were performed in a
14-m-long and 3-m-wide corridor. Each pedestrian stream entered this
corridor from one of the two ends of the corridor. The dimensions
and configuration of the region of interest (ROI) are shown in
figure \ref{fig1}. Let $\rho^c$ and $v^c$ be the average density and
speed of the reference stream, respectively. Here, the average
density $\rho^c$ (in ped/m$^2$) and the average speed $v^c$ (in
m/s) in the ROI are computed as the number of reference people
in the ROI divided by the area of the ROI and the sum of their
walking speeds divided by the area of the ROI, respectively. The
specific flow of the reference stream is expressed as $J^c$ (in
ped/m/s), which is equal to $\rho^c\times v^c$.

\begin{figure}
\begin{center}
\includegraphics[width=\textwidth]{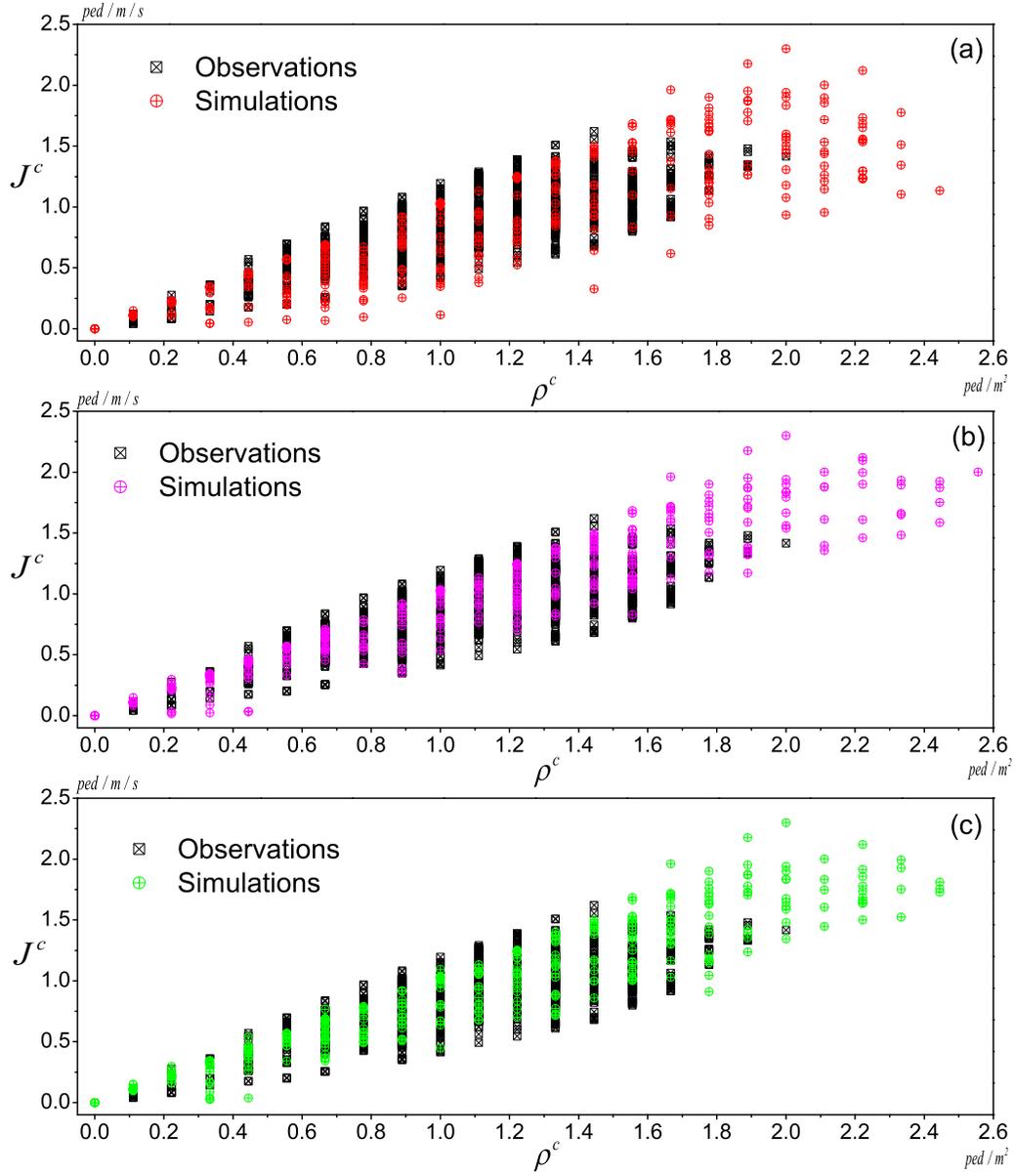}
\caption{\label{fig2}Comparisons of fundamental diagrams of
observations and simulation data. (a) $\omega_1=0$;
(b) $\omega_1=0.1$; (c) $\omega_1=0.2$.}
\end{center}
\end{figure}

Figure \ref{fig2} compares the fundamental diagrams of bidirectional
pedestrian flow for the camera-based observations and simulation
results obtained with different values of the weight coefficient
$\omega_1$.  There are no extra collision conflicts between the two
streams at $\omega_1=0$. At larger $\omega_1$ values, pedestrians
have a more obvious tendency to reduce collision conflicts with
other pedestrians in the conflicting stream. Figure \ref{fig2} shows
that the fundamental diagrams for the simulation data obtained with
various values of $\omega_1$ agree well with those for the
observations, and no large differences are found in the observed
density range, $\rho^c<2$~ped/m$^2$. The specific flow $J^c$
increases as the density $\rho^c$ increases from 0 to $2$~ped/m$^2$,
which implies that the reference stream is in an uncongested state.
Differences in the fundamental diagrams for the simulation data can
be observed for densities $\rho^c>2$~ped/m$^2$. From figure
\ref{2}(a), the specific flow $J^c$ for $\omega_1=0$ clearly
decreases as the density $\rho^c$ increases beyond $2$~ped/m$^2$,
which indicates that the reference stream is in a congested state.
Therefore, the critical density at which the maximum flow is
achieved is about $2$~ped/m$^2$ for $\omega_1=0$. The critical
density for $\omega_1>0$ is slightly larger than that for
$\omega_1=0$, indicating that the probability of congestion will
decrease at a larger value of $\omega_1$ under the same conditions.
The credibility of the data produced by the extended model can be
validated by comparing them with the findings from the controlled
experiments of bidirectional pedestrian streams.

\begin{figure}
\begin{center}
\includegraphics[width=\textwidth]{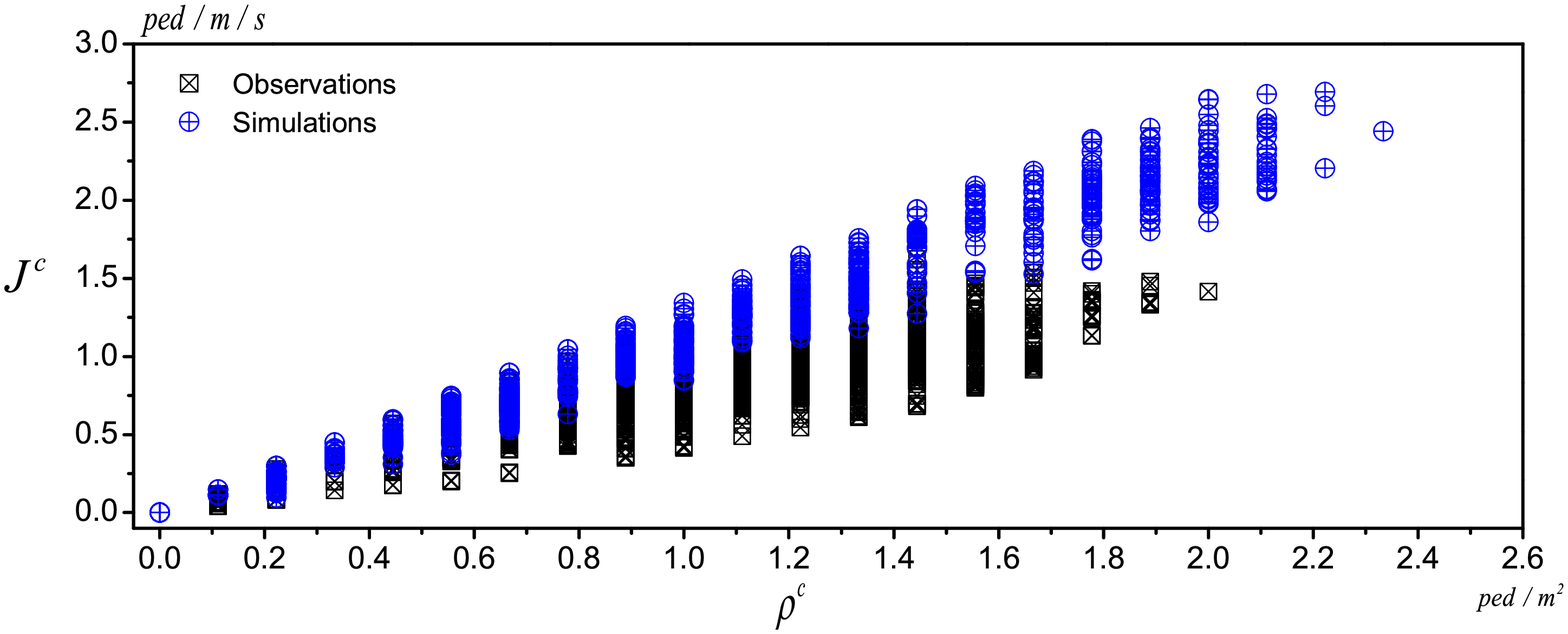}
\caption{\label{fig3}Comparisons of fundamental diagrams of
observations and simulation data obtained using Helbing's model.}
\end{center}
\end{figure}

Figure \ref{fig3} compares the fundamental diagrams of bidirectional
pedestrian flow for the camera-based observations and the simulation
results obtained by Helbing's model. In this model, the desired
walking directions for the two streams are set to $(1,0)$ and
$(-1,0)$, respectively. The specific flow $J^c$ based on the
previous experimental results is found to be significantly larger
than that based on Helbing's model, especially at relatively high
density (i.e., $\rho^c>0.7$~ped/m$^2$). Thus, the setting of the
desired directions in Helbing's model, which describes the
shortest-path strategy of pedestrians, is inappropriate at
relatively high density. In contrast, figure \ref{fig2} shows that
the reactive user-equilibrium path choice strategy introduced in the
extended model is rather appropriate in this case.

Using open boundary conditions, we conduct numerical simulations of
bidirectional pedestrian movement in a $40$~m$\times10$~m corridor.
Initially, the corridor is empty. Pedestrians enter the corridor
from both ends at random. The inflow value for each group is set to
$6$~ped/s during a simulation that lasts $600$~s, and the inflow
pattern is similar for all simulations. Note that $600$~s for each
simulation is long enough for the bidirectional flow study because
the length of the corridor in our simulation is only $40$~m, and a
pedestrian requires only about $40$~s to traverse the corridor.

\begin{figure}
\begin{center}
\includegraphics[width=\textwidth]{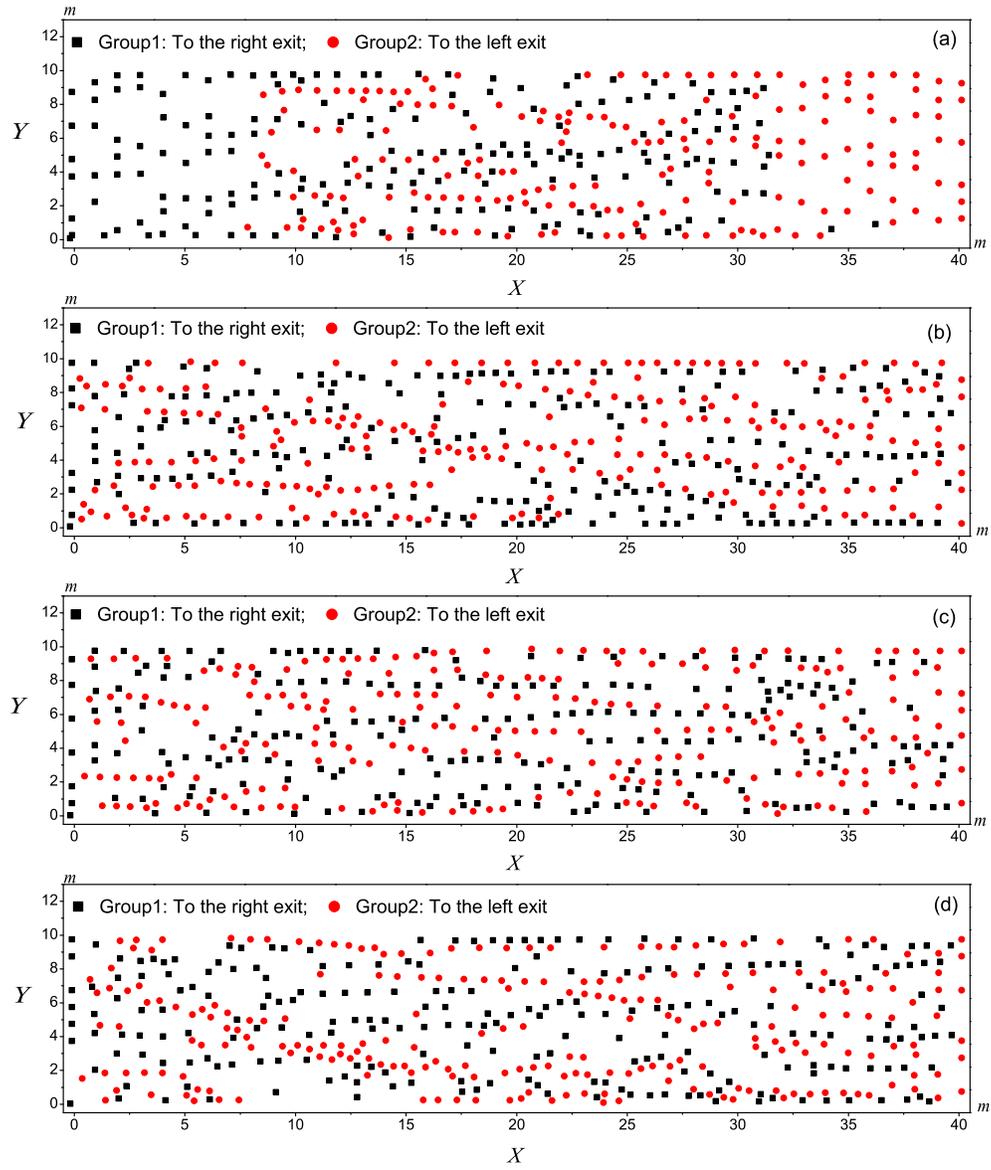}
\caption{\label{fig4}Spatial distribution of pedestrians obtained
with $\omega_1=0$. (a) $t=30$~s; (b) $t=40$~s; (c) $t=240$~s;
(d) $t=480$~s.}
\end{center}
\end{figure}

Figure \ref{fig4} shows snapshots of the simulation process obtained
with the weight coefficient $\omega_1=0$. The black
squares represent pedestrians walking toward the right exit of the
corridor, and the red balls represent pedestrians walking toward the
left exit. We can see
clearly that the opposing flows are mixed, and unstable lanes
consisting of pedestrians who intend to walk in the same direction
are formed. The lanes vary in time and space. This regime is
the dynamic multilane (DML) regime observed in controlled experiments of
bidirectional pedestrian movement \cite{Zhang2012}.

\begin{figure}
\begin{center}
\includegraphics[width=\textwidth]{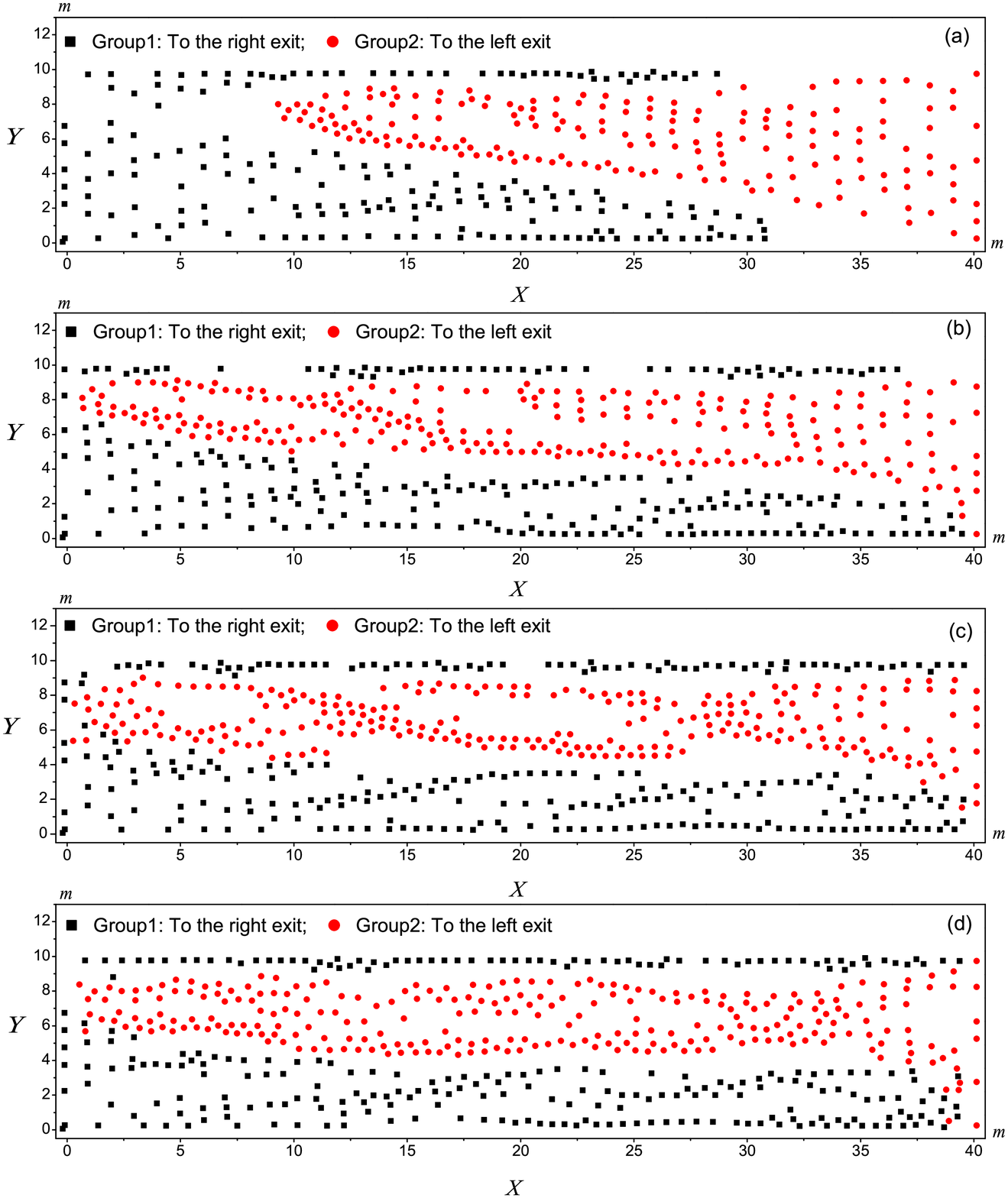}
\caption{\label{fig5}Spatial distribution of pedestrians obtained
with $\omega_1=0.1$. (a) $t=30$~s; (b) $t=40$~s; (c) $t=240$~s;
(d) $t=480$~s.}
\end{center}
\end{figure}

Figure \ref{fig5} shows snapshots of the simulation process
obtained with $\omega_1=0.1$. We observe
complete segregation of the two opposing streams as a result of the
two groups of pedestrians' strong collision avoidance behavior. This
self-organized pattern of motion enhances the pedestrian flow by
reducing the collision probability. Three separate lanes
are clearly formed autonomously immediately after the
simulation begins and remain stable during the simulation process. This
regime is called stable separate lanes (SSL) regime \cite{Zhang2012}.

\begin{figure}
\begin{center}
\includegraphics[width=\textwidth]{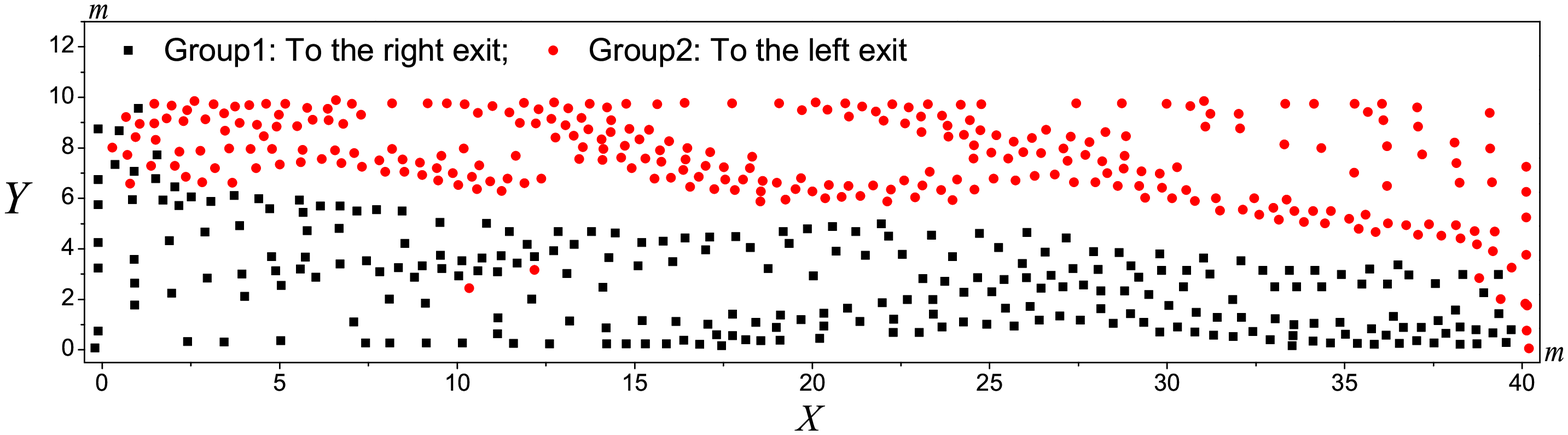}
\caption{\label{fig6}Spatial distribution of pedestrians obtained
with $\omega_1=0.2$ at $t=480$~s.}
\end{center}
\end{figure}

\begin{figure}
\begin{center}
\includegraphics[width=\textwidth]{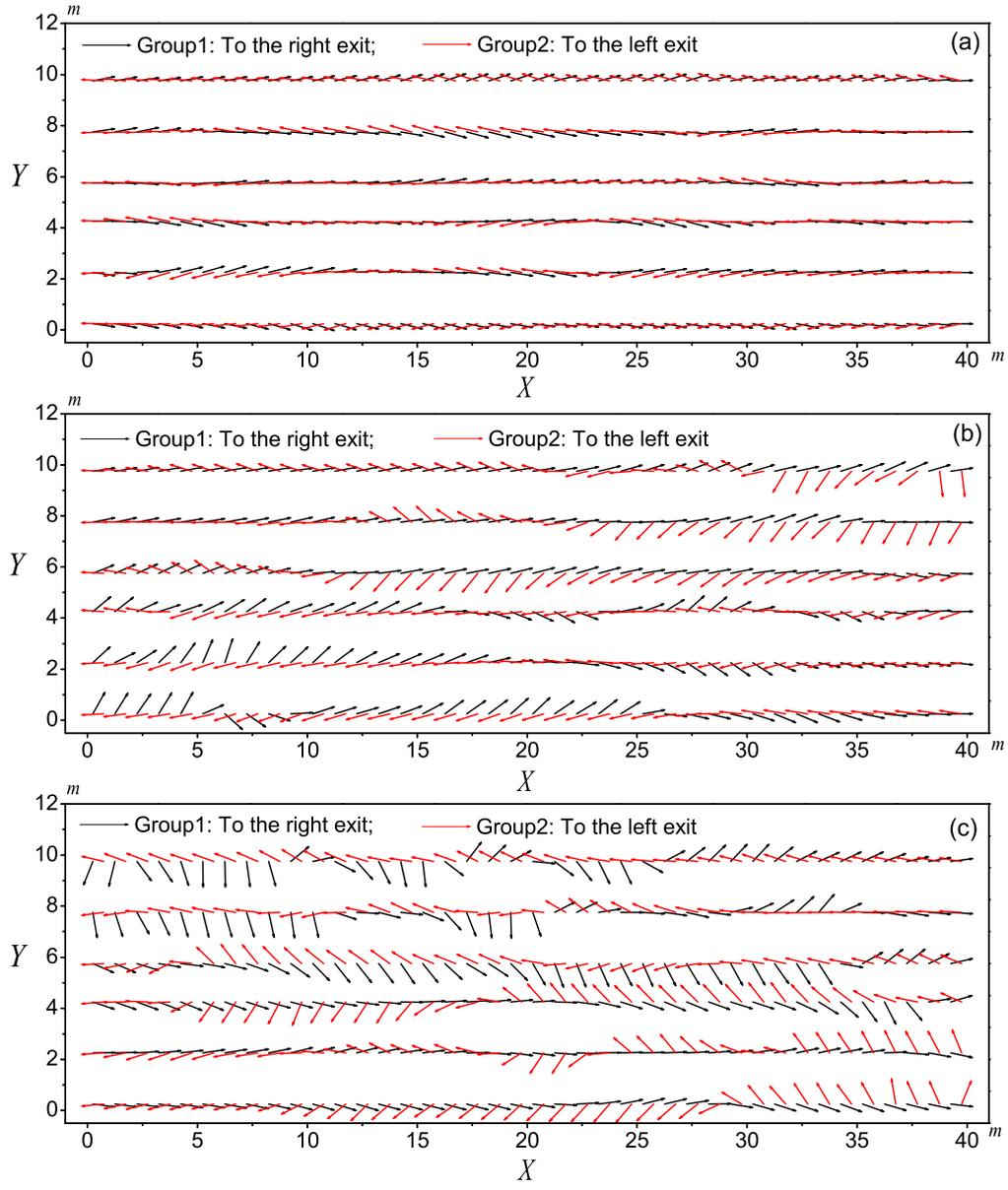}
\caption{\label{fig7} Dynamic navigation fields at $t=480$~s.
(a) $\omega_1=0$; (b) $\omega_2=0.1$; (c) $\omega_1=0.2$.}
\end{center}
\end{figure}

SSL flows are also observed for $\omega_1=0.2$ (figure \ref{6}).
From figure (\ref{6}), the opposing streams segregate themselves and occupy
identical shares of the corridor, as if there is a partition line in
the corridor; i.e., a quasi-symmetric spatial distribution is
formed. The number of self-organized lanes decreases, as pedestrians
have a more obvious tendency to reduce collision conflicts with other
pedestrians in the conflicting stream. It is observed that the
optimal path choice strategy of pedestrians can cause different
spatial distributions of pedestrians in the corridor according to the
value of the weight coefficient $\omega_1$.

Figure \ref{fig7} plots the dynamic navigation fields obtained for
$\omega_1=0, 0.1$, and $0.2$ at time $t=480$~s. The dynamic
navigation fields describe the desired directions of two groups of
pedestrians and direct pedestrians to move toward the exits of the
corridor. The navigation fields are time-varying according to the
instantaneous pedestrian distribution in the corridor. They are
generated by the optimal path choice strategy of pedestrians who
intend to choose a path with the lowest instantaneous walking cost.
This figure shows that the probability of face-to-face collisions
between two pedestrians in bidirectional pedestrian flow is
significantly reduced as $\omega_1$ increases. This path choice
strategy also produces various self-organized patterns of
bidirectional pedestrian movement (figures \ref{fig4}--\ref{fig6}).

\section{Conclusion}
\label{sec:level5} Bidirectional pedestrian motion is
studied using an extended social force model with a dynamic
navigation field. The dynamic navigation field, which arises from the
decision-making processes of pedestrians, is introduced to describe
the desired direction of bidirectional pedestrian motion. Here, the
desired walking direction of each group is described so as to minimize the total
instantaneous cost in a reactive user-optimal manner and to reduce
collision conflicts with other pedestrians in the conflicting group.
The extended model of bidirectional pedestrian flow is calibrated by
comparing empirical observations from controlled experiments and
simulation data obtained using the simulation model. Numerical results
show that pedestrians' path choice behavior significantly affects
the self-organized patterns of bidirectional pedestrian movement,
e.g., the DML and SSL patterns. The
probability of congestion generally decreases with increasing tendency to reduce collision conflicts. Further, the number of
self-organized lanes also decreases as the tendency to reduce
collision conflicts increases.

\vspace{.2in}

\noindent{\bf Acknowledgments} \vspace{.1in}

This work was supported by the National Natural Science Foundation
of China (Grant Nos. 11202175, 11275186, 91024026, and FOM2014OF001),
the Research Foundation of Southwest University of Science and
Technology (No. 10zx7137), and a Singapore Ministry of Education
Research Grant (Grant No. MOE 2013-T2-2-033).

\end{document}